\documentclass[
    ,final            
  ]
  {aipproc}

\layoutstyle{6x9}

\usepackage{amsmath}
\usepackage{color}

\begin{document}

\newcommand{\xxx}[1]{\textbf{\sffamily \color{red}  #1}}

\title{A new extraction of neutron structure functions from existing inclusive DIS data}


\author{J. G. Rubin}{
  address={Physics Division, Argonne National Laboratory, Argonne, Illinois 60439, USA}
}

\author{J. Arrington}{
  address={Physics Division, Argonne National Laboratory, Argonne, Illinois 60439, USA}
}

\begin{abstract}

A recent reanalysis of world proton and deuteron structure function
measurements showed that a significant amount of the apparent model dependence
in the extraction of the neutron structure function was related to
inconsistencies between the kinematics of the data and those assumed for the
calculation, suggesting that the true model dependence is smaller than
commonly believed.  We present a detailed comparison of the neutron structure
function as extracted using different models, with care taken to ensure that
all other aspects of the comparison are done consistently.   The neutron
structure function is extracted using a fit to these data evaluated at fixed
$Q^2_0$=16~GeV$^2$.  We compare the results obtained using a variety of N--N
potentials and deuteron binding models to determine the model dependence of
the extraction.  As in the recent extraction, $F_{2n}/F_{2p}$ falls with $x$
with no sign of plateau and follows the low edge of the wide range of earlier
$F_{2n}$ extractions.  The model-dependent uncertainty in $F_{2n}/F_{2p}$ is
shown to be considerably smaller than previously believed, particularly at
large-$x$.

\end{abstract}

\maketitle


\section{Introduction}

\begin{figure}[htb]
\includegraphics[width=.75\textwidth]{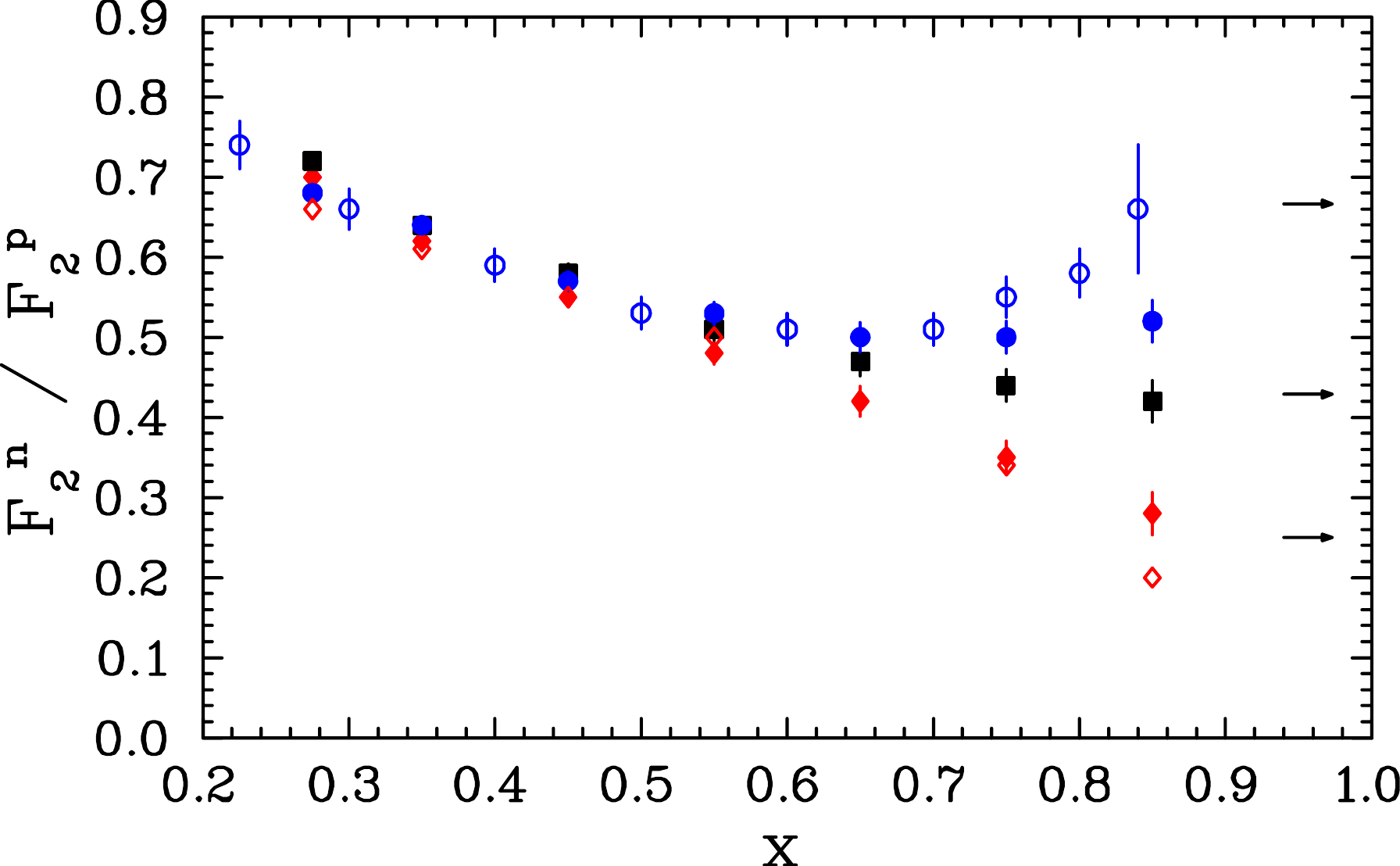}
\caption{Several prior extractions of the ratio of the neutron to proton
structure functions showing a large range, particularly at high-$x$.
The models shown are: off-shell (black squares) and on-shell (solid
red diamonds) extractions from Ref.~\cite{melnitchouk96}, pure Fermi-motion
(open red diamonds)~\cite{Whitlow:1991uw}, and 
nuclear density dependent EMC effect based models (solid and open blue
circles) from Ref.~\cite{Whitlow:1991uw} and \cite{frankfurt88} respectively. 
The arrows on the right-hand side of the figure indicate various theoretical
predictions of this ratio in the $x=1$ limit~\cite{melnitchouk96}.  Exact
spin-flavor SU(6) symmetry implies a value of 2/3, symmetry breaking through
vector ($S=1$) diquark suppression at large-$x$ gives a value of 1/4, and a
value of 3/7 is obtained if only the $z$-component, $S_z=1$ of the vector
diquark's spin determines its suppression.}
\label{fig:oldExtractions}
\end{figure}

Because of the experimental impracticality of performing precision scattering
experiments with a neutron target, the determination of the structure
functions of the neutron requires 1) inclusive DIS scattering data from both
proton and deuteron targets and 2) a theoretical model that convincingly
describes the impact on the structure function of the proton and neutron when
bound together.  Previous extractions of the neutron structure function, and
more relevantly the ratio of the structure function of the neutron to that of
the proton, $F_{2n}/F_{2p}$, were performed using different subsets of the
world data on $F_{2p}$ and $F_{2d}$ with varying treatments of binding and
kinematics.  These produced significant variation in the resulting ratio,
particularly at large-$x$.  Some of these are shown in Figure
\ref{fig:oldExtractions}.  Because these different approaches yielded very
different results, it was concluded that the model dependence of such an
analysis was so large that no reliable information on the neutron could be
obtained at large-$x$.  However, as discussed in~\cite{arrington09}, a
significant amount of the variation between results is related to
self-inconsistencies in the extractions, giving the impression that our
knowledge of this quantity is significantly worse than it will actually be
shown to be.  By carefully and uniformly treating the source data and
restricting deuteron models to a set that is modern and compelling, a
reasonable and significantly smaller uncertainty estimate can be made.

\section{Extracting $F_{2n}/F_{2p}$}

\begin{figure}[htb]
\includegraphics[width=.47\textwidth]{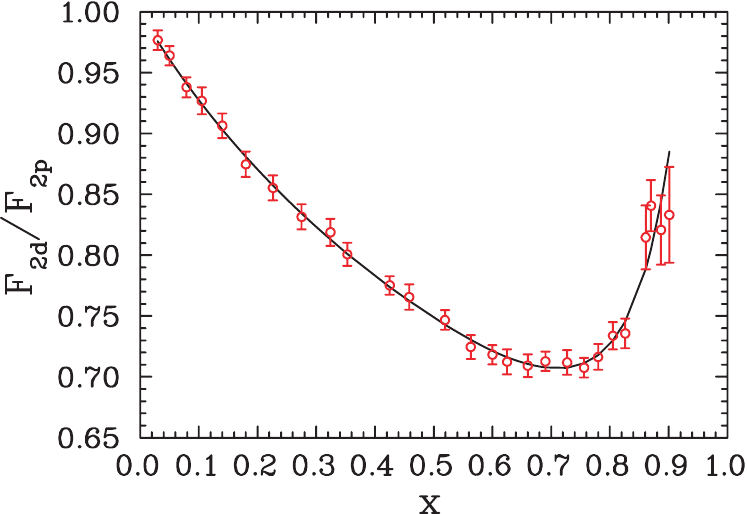}
~~~~
\includegraphics[width=.46\textwidth]{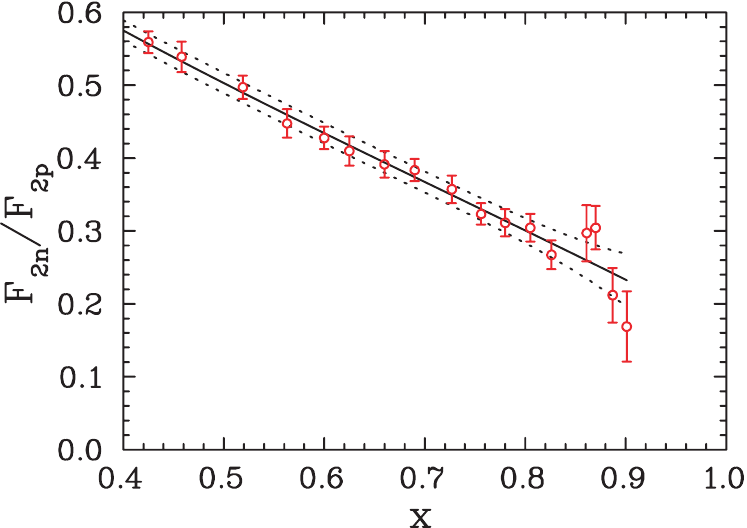}
\caption{Left panel: $F_{2d}/F_{2p}$ data from the range 6 < $Q^2$ < 30~GeV
interpolated to a common $Q^2_0$=12~GeV$^2$ shown with a parameterization. 
Right panel: The extracted parameterization of $F_{2n}/F_{2p}$ with normalization
uncertainty bands propagated from the proton and deuteron data sets.  The data
points are generated from the original $F_{2n}/F_{2p}$ values and the smearing
ratios produced using the $R_{np}$ parameterization.  Figures reproduced from
\cite{arrington09}.}
\label{fig:newRatios}
\end{figure}

We begin with a summary of the extraction performed in Ref~\cite{arrington09},
as this provides the starting point for this comparison.  The proton input,
$F_{2p}(x,Q^2)$, used in this analysis comes from a parameterization provided
by M. E. Christy, which is fit to a large body of experimental experimental
results.  The deuteron to proton structure function ratio, $F_{2d}/F_{2p}$, is
taken from SLAC, BCDMS, and NMC measurements covering a range in $Q^2$ from 3
to 230~GeV$^2$. Data points for which $6 < Q^2 < 30$~GeV$^2$ and $W^2 >
$3.5~GeV$^2$ are interpolated to a common $Q_0^2$ of 12~GeV$^2$ (typically a
0.5\% correction).  The data points and fit function are shown in the left
panel of Figure~\ref{fig:newRatios}.	

The nuclear modification of the proton and neutron was computed in a light
cone impulse approximation, without any assumption of scaling or neglecting
$k_\perp$, using the CD-Bonn~\cite{PhysRevC.63.024001} effective two-body
potential.  This model is applied to the proton data and a $R_{np}$~($\equiv
F_{2n}/F_{2p}$) ratio of the form
\begin{equation*}
R_{np}(\xi,Q^2) = (p_1 + p_2~\xi) + p_3~e^{-p_4~\xi} + p_5~e^{-p_6~(1-\xi)}+p_7~\textrm{max}(0,\xi-p_8)^2,
\end{equation*}
where $\xi = 2x/(1+\sqrt{1+4M^2x^2/Q^2})$ is chosen as the scaling variable to
reduce the $Q^2$ dependence.  The free parameters are tuned using
\texttt{MINUIT} to best reproduce the $F_{2d}$ data in conjunction with the
parameterized proton structure function.  The result of this procedure is the
curve in the right panel of Figure~\ref{fig:newRatios}.  
Once $R_{np}$ is determined, smearing ratios $S_p=\tilde{F}_{2p}/F_{2p}$ and
$S_n=\tilde{F}_{2n}/F_{2n}$, where the tilde indicates the structure function
of the bound nucleon, are trivial to determine.  Using these ratios, the
individual $R_{dp}=F_{2d}/F_{2p}$ points can be converted to $R_{np}$ values,
along with their individual uncertainties.  In addition, detailed estimates of
the systematic uncertainties associated with the extraction procedure and data
normalization have been parameterized.  The individual $R_{np}$ points as well
as the systematic uncertainties are shown as the points and dotted bands in
the right panel of Figure~\ref{fig:newRatios}. 

\begin{figure}[htb]
\includegraphics[width=.75\textwidth]{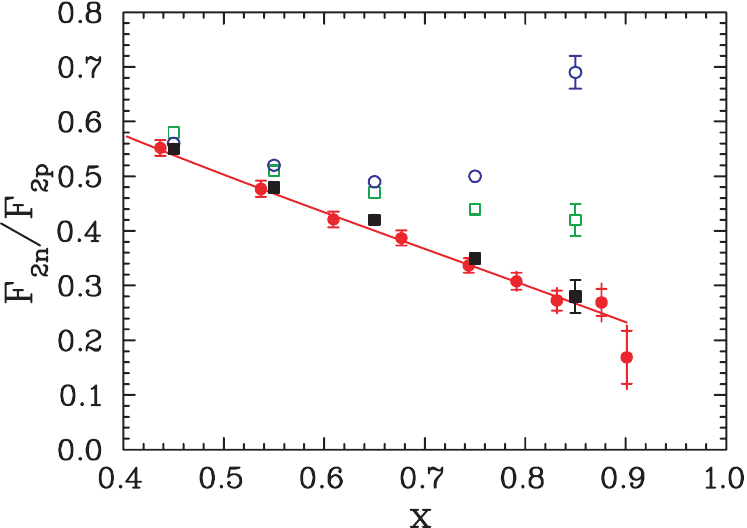}
\caption{The resulting parameterization (red line) and data points (solid red
circles) compared with previous calculations.  An off-shell model
\cite{melnitchouk96} (open green squares), on-shell model
\cite{PhysRevD.49.4348} (solid black squares), and a modified EMC effect model
(open blue circles)~\cite{Whitlow:1991uw} are shown.  Reproduced from
\cite{arrington09}.}
\label{fig:extractionCompWithOld}
\end{figure}

Figure~\ref{fig:extractionCompWithOld} compares the results of this extraction
to previous results, with the new data points rebinned to approximately match
the previous binning.  These results fall on the lower edge of the wide range
of previous results.  Note that this extraction was performed using a single
model of the deuteron structure and the CD-Bonn N--N potential.  No estimate
of the model dependence due to choice of nuclear models was included.

\begin{figure}
\includegraphics[width=.41\textwidth]{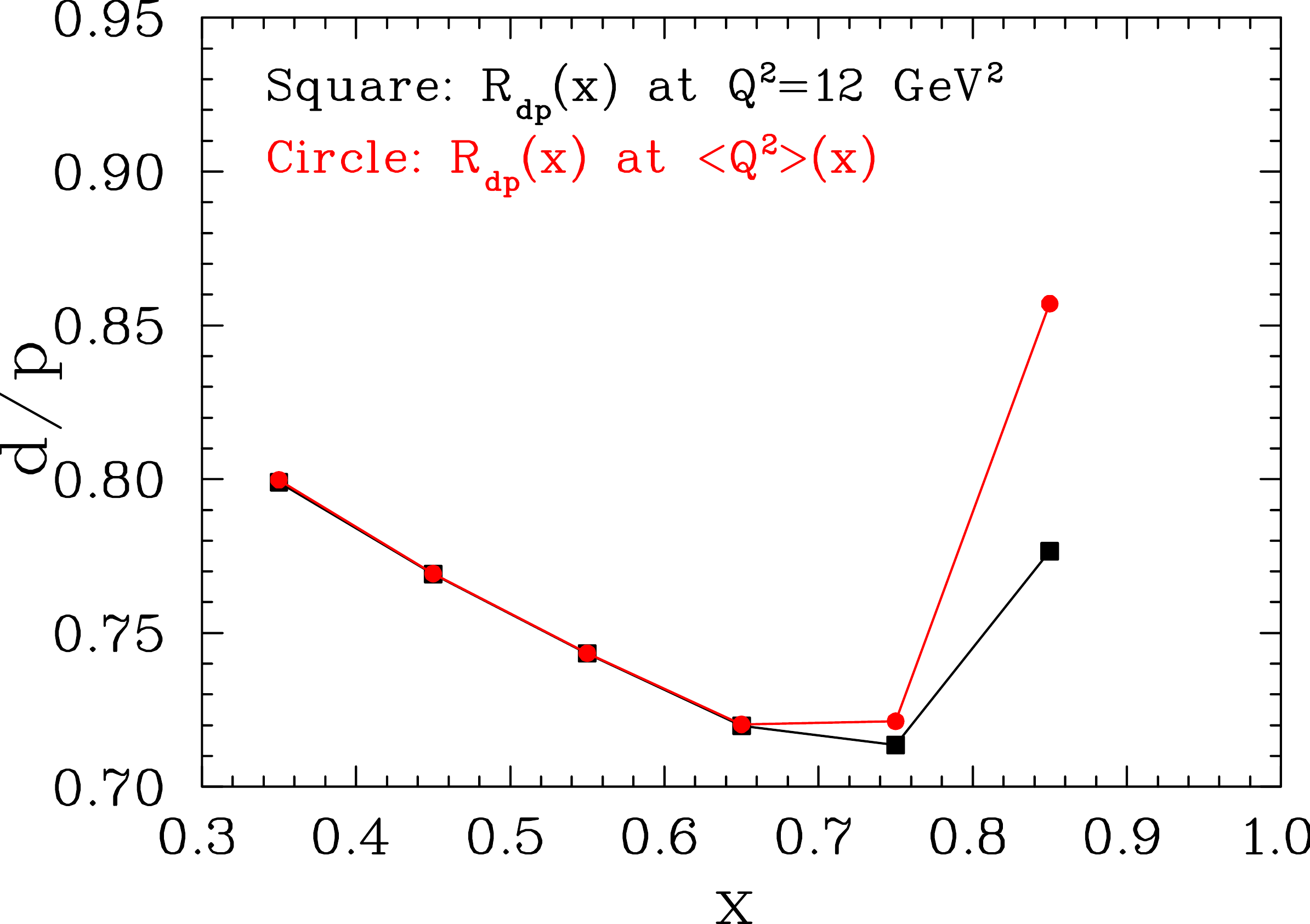}
~~~~
\includegraphics[width=.54\textwidth]{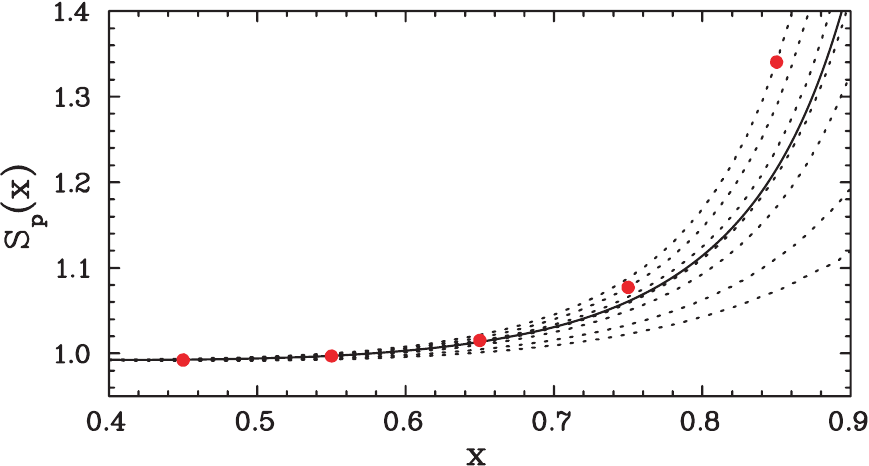}
\label{fig:Q2}
\caption{Left panel: The calculated ratio $F_{2d}/F_{2p}$ evaluated at the
average $Q^2$ for the given $x$ value (red), and at a common $Q^2_0$ =
12~GeV$^2$ (black).  Right panel: The proton smearing ratio, $S_p$ for values
of $Q^2$ ranging from 4.7 to 23.6~GeV$^2$ (corresponding to the average
$Q^2$ value of each of the $R_{dp}$ points).  The dark central line corresponds
to the $Q^2_0$=12~GeV$^2$ used in this analysis.  The red points indicate 
the value of $S_p$ at the average $Q^2$ of the input data.}
\end{figure}

Most previous extractions of the neutron structure function treated the data
as though it were at a fixed $Q^2$ when in fact there is a strong correlation
between $x$ and $Q^2$ as one can see in the left panel of Figure~\ref{fig:Q2}.
Furthermore, even the simple convolution models including only Fermi motion
yield a significant $Q^2$ dependence at large x.   This can be seen for
the calculation of the smearing ratio in right panel of Figure~\ref{fig:Q2}
where $S_p(x)$ is given for different values of $Q^2$ and, for comparison,
$S_p(x)$ points at the $Q^2$ of the data are overlaid. Neglecting this
$x$-$Q^2$ correlation by applying nuclear corrections at fixed $Q^2$ can yield
a significant error in the extraction.   There are two ways to address this
correlation.  The first is to evolve the model of the deuteron to the $Q^2$
value of each data point, which is essentially the procedure used in the CTEQ6X
global extraction of parton distributions~\cite{accardi10}. The
second, which is used in this work is to interpolate all of the data to a
fixed value of $Q^2$.  A benefit of this method is that the analysis can be
performed purely in terms of structure functions, without invoking the parton
picture, which leads to a particularly direct and accessible analysis scheme.
Both of these recent approaches yield a similar result, with $R_{np}$
approaching 1/4 (and thus the $d(x)/u(x)$ ratio approaching zero) as $x$
becomes very large. 	

\section{Exploring Deuteron Model-Dependence}

Given the observation that accounting for $Q^2$-dependent contributions is
important in these analyses~\cite{arrington09,accardi10}, we perform a
detailed study of the model dependence of the extracted neutron structure
function when fully and consistently accounting for these effects in the
comparison.  We factor this problem into two parts: 1) the choice of a N-N
potential and 2) the choice of a nuclear model.  We begin by updating the
interpolation of the global $R_{dp}$ measurements to fixed $Q^2$, this time
choosing $Q^2$=16~GeV$^2$, as this more closely matches the kinematics of the
large-$x$ measurements, where the $Q^2$ dependence is largest.  While this
should have no impact on the comparison of the different extractions, it
should minimize the corrections in the interpolation for all of the results.

\begin{figure}[htb]
\includegraphics[width=.49\textwidth]{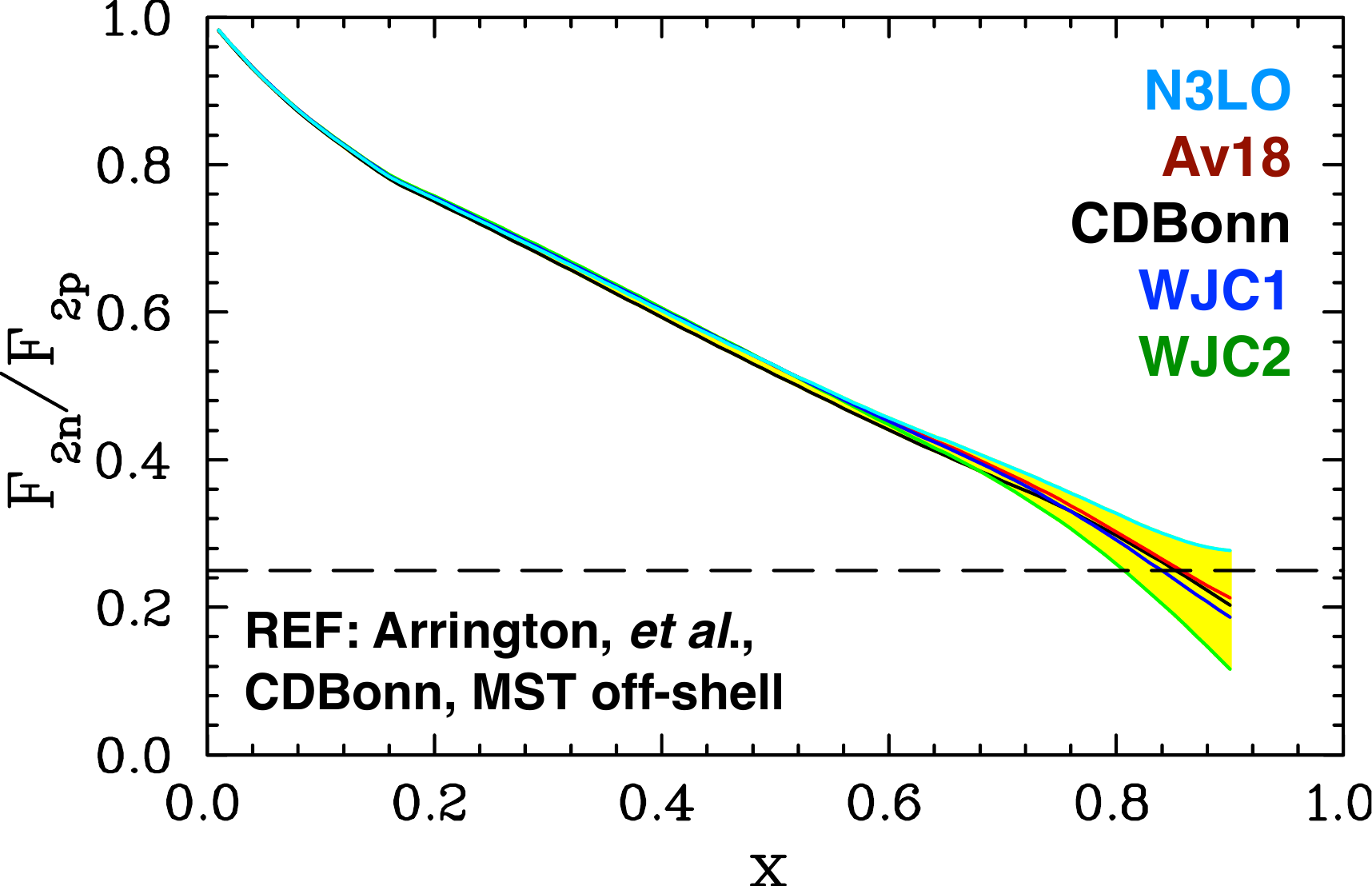}
\includegraphics[width=.49\textwidth]{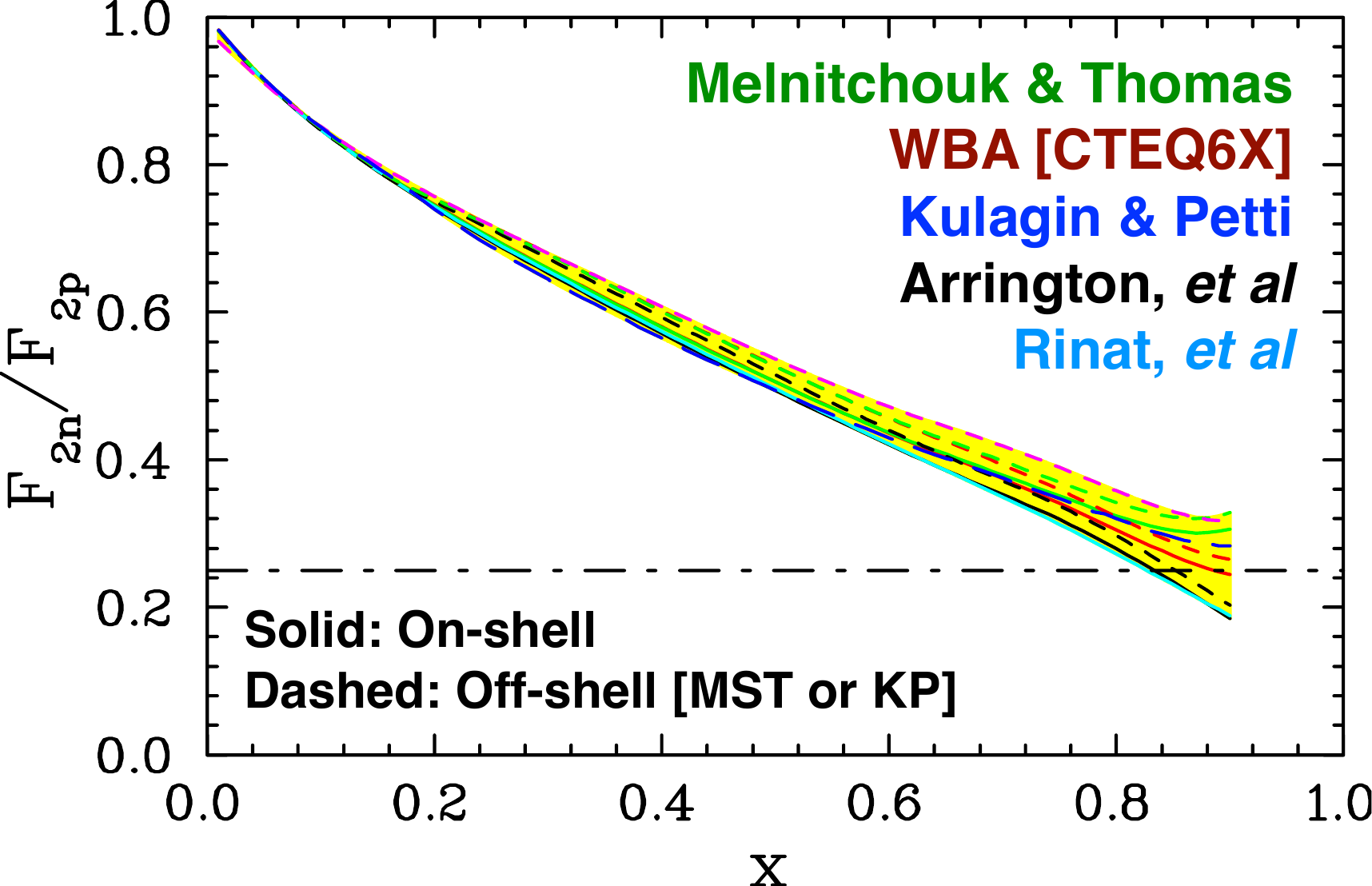}
\caption{Left panel: $R_{np}$ calculated using various N-N potentials assuming
the deuteron model described previously~\cite{arrington09}. Right panel:
$R_{np}$ calculated using different models~\cite{melnitchouk96, kulagin06,
rinat07, kahn09, arrington09} with the CD-Bonn N-N potential
\cite{PhysRevC.63.024001}.}
\label{fig:modelDep}
\end{figure}

We extract $R_{np}$, repeating the analysis of Ref.~\cite{arrington09} at
$Q_0^2$=16~GeV$^2$.  We then take our input fit to $F_{2p}$ along with the
extracted $F_{2n}$ as the starting point to calculate the smearing ratios
(ratio of the proton or neutron structure function in the deuteron to the free
structure function) for all of the different models, and use this to determine
the change in the extracted neutron structure function. Variations in $R_{np}$
can be studied for potential and model separately by varying the potential
with a single baseline model and varying the model with a single baseline
potential.  These two procedures are depicted in the left and right panels of
Figure~\ref{fig:modelDep} respectively.  The model of Ref.~\cite{arrington09}
is used when comparing the different N--N potentials, while the CD-Bonn
potential is used for comparison of the different calculations of the nuclear
effects.  We take the full range of the results shown in each panel as the
one-sigma systematic band for the potential and model dependences of the
extraction.

Note that in this comparison, we use models which include Fermi
motion, binding, and in some cases off-shell effects and the contributions of
nuclear pions.  All of these represent ``ground-up'' calculations of the
deuteron structure based on input proton and neutron structure functions and
some N--N potential used to calculate the deuteron momentum distribution. 
Some of the earlier models which yielded exceptionally large results for
$R_{np}$ at large-$x$ were based on models which neglected Fermi motion and
binding and yielded a large effect through significant modification of the
nucleon structure in the nucleus, e.g. the ``scaled EMC effect'' result
of~\cite{Whitlow:1991uw}.  While such large nuclear effects are possible, they
are not implemented in a realistic fashion in this approach.  They do not
include any direct calculation of Fermi motion, are scaled down for the deuteron
using an extremely large assumed nuclear density density, neglect $Q^2$
dependence in the nuclear effects, and effectively assume $S_n=S_p$ which is
not true when there is a significant difference in the $x$ dependence
of $F_{2p}$ and $F_{2n}$.  Thus, we do not include models that
use an explicit ``EMC effect'' for the deuteron, unless they are
included (e.g. via the pion contributions and off-shell effect) 
along with Fermi motion and binding fashion~\cite{melnitchouk96, kulagin06,
kahn09, accardi10}.

We note that the issue of the $Q^2$ variation of the $R_{dp}$ measurements is
important even for the models including a large EMC effect in the deuteron.
The model used to obtain the highest set of data points from
Figure~\ref{fig:oldExtractions} yields a smaller neutron structure function at
large $x$ when applied to the $R_{dp}$ measurements interpolated to a fixed
$Q^2$ value.  Thus, even if one were to include such models, the impact on the
neutron structure function is smaller than one would expect based on these
earlier comparisons.

\begin{figure}
\includegraphics[width=.75\textwidth]{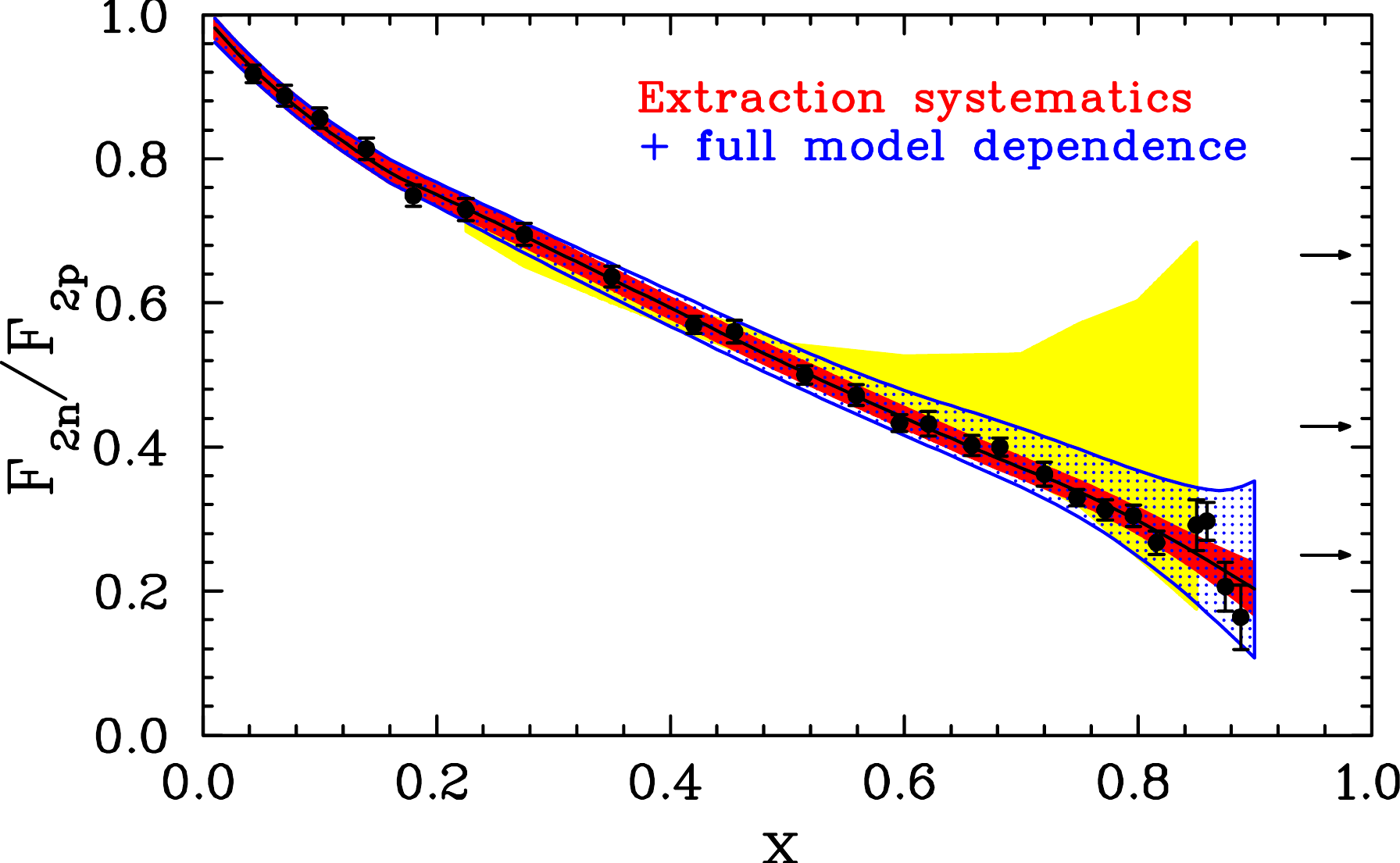}
\caption{The extracted $F_{2n}/F_{2p}$ with the propagated normalization
systematic uncertainty band in red and the total deuteron theory uncertainty
in blue.  The yellow band represents the range of prior extractions of this
quantity.  The arrows on the right once again indicate various values in the
$x$=1 limit under certain theoretical assumptions as described in Figure
\ref{fig:oldExtractions}.}
\label{fig:resultWithOldSets}
\end{figure}

\begin{figure}
\includegraphics[width=.95\textwidth]{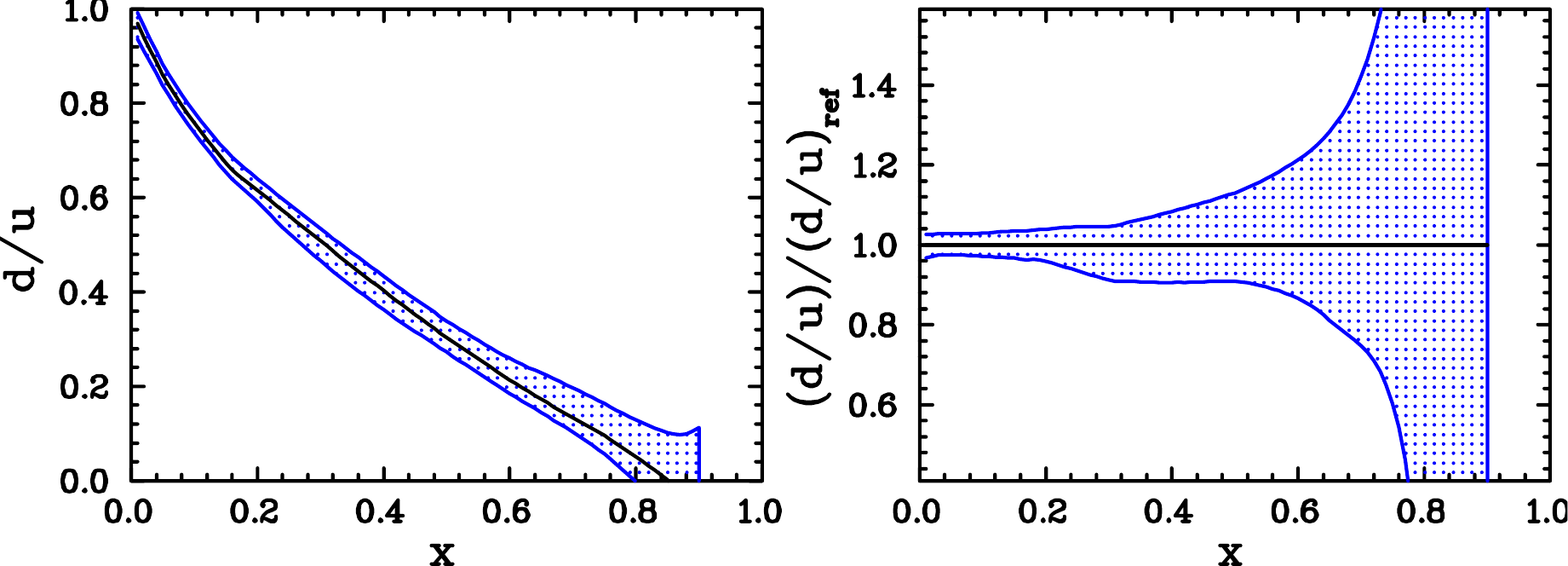}
\caption{The extracted $d/u$ ratio for the proton with extracted error
band.  The left panel shows the absolute value, while the right panel shows
the band relative to the central value.}
\label{fig:dtou}
\end{figure}

Finally then, the systematic uncertainty bands from the experimental
extraction (as parameterized in Ref.~\cite{arrington09}) and the model and
potential dependence of the result are combined in quadrature.  This is shown
in Figure~\ref{fig:resultWithOldSets} compared to the much larger range of
prior $F_{2n}/F_{2p}$ extractions.  Based on these results, it appears that
the neutron structure function is relatively well known even to large $x$
values, under the assumption that there are no ``exotic'' contributions to
the deuteron structure beyond the effects included in these models.  This
conclusion may seem to be at odds with the CTEQ6X analysis~\cite{accardi10}
where, even with the inclusion of large-$x$ data, the uncertainties in the
high-$x$ $d$-quark distributions are extremely large.  Figure~\ref{fig:dtou}
shows our extracted result for the value and uncertainty in the $d/u$ ratio
for the proton, where we extract this from $F_{2n}/F_{2p}$ by neglecting
contributions from strange and heavier quarks.  The left panel shows the
absolute value of the $d/u$ ratio, and it is clear from this that the
$d$-quark distribution is becoming very small.  The right panel shows the
ratio compared to the reference fit, and shows that the relatively small
absolute uncertainties on $F_{2n}/F_{2p}$ or $d/u$ can yield $>$100\%
uncertainties on the absolute value of the $d$-quark distributions.  Thus, the
uncertainty on the $d$-quark distribution is small relative to the size of the
dominant $u$-quark distribution, which means that the data can yield
significant constraints when comparing to predictions of the $d/u$ ratio at
large $x$. However, the \textit{fractional} uncertainty on the $d$-quark
distribution is large, meaning that observables that are sensitive to the
$d$-quark pdf at large $x$ are not well constrained.  So the impact of the
nuclear models is small for some purposes but large in others, and the context
is clearly important in determining whether we have sufficiently precise
knowledge of the neutron structure function.

There are multiple model-independent extractions planned for the neutron
structure function at large-$x$ as part of the 12~GeV upgrade
plan~\cite{other_talks}. These will improve the precision of the neutron
extraction at high $x$, but more importantly will be sensitive to any physics
beyond what is contained in these models, and thus are very sensitive to some
of the more exotic explanations of the EMC effect.

\section{Conclusions}

A recent reanalysis of inclusive DIS data from proton and deuteron scattering,
interpolating to a common value of $Q^2$ provided a new extraction of 
$F_{2n}/F_{2p}$.  The result was found to decrease as $x$ becomes large and
shows no sign of plateau, sitting near the low end of the wide range of
previously extracted large-$x$ behaviors.  However, some previous extractions
yielded a neutron structure function that was too large because of
inconsistent treatment of $x$-$Q^2$ correlations.  We have extended this
analysis to include a detailed extraction of the model dependence, using a
range of deuteron models and potentials.  We find that the sensitivity of the
extracted $F_{2n}/F_{2p}$, and hence the model uncertainty, is relatively
small, and significantly less than previously believed.  While these results
could be an underestimate if there is a larger than expected ``EMC effect'' in
the deuteron, the comparison of this extraction with future model-independent
measurements will be able to measure or significantly limit such modification
of the nucleon structure in the deuteron.

\begin{theacknowledgments}

The authors would like to thank Roy Holt for initiating this work and Wally
Melnitchouk for providing many of the calculations that are crucial in this
analysis. This work is supported by the U.S. Department of Energy, Office of
Nuclear Physics, under contract DE-AC02-06CH11357.
\end{theacknowledgments}


\bibliographystyle{aipproc}   

\bibliography{neutron}

\end{document}